\documentclass[abstract=true, DIV=16, parskip=half]{scrartcl}

\usepackage[noEnd,commentColor=black]{algpseudocodex}
\usepackage{amsmath}
\usepackage{amssymb}
\usepackage{amsthm}
\usepackage[mathscr]{euscript}
\usepackage[shortlabels]{enumitem}
\usepackage{graphicx} 
\usepackage{subcaption}
\usepackage{thmtools}
\usepackage{todonotes}
\usepackage{xcolor}
\usepackage{microtype}

\makeatletter
\def\@fnsymbol#1{\ensuremath{\ifcase#1\or \!\;\or \!\;\or \ddagger\or
   \mathsection\or \mathparagraph\or \|\or **\or \dagger\dagger
   \or \ddagger\ddagger \else\@ctrerr\fi}}

\makeatother

\usepackage[normalem]{ulem}
\usepackage{tcolorbox}

\usepackage{upgreek}
\usepackage[colorlinks]{hyperref}
\usepackage[capitalize]{cleveref}

\crefname{equation}{eq.}{eqs.} 
\crefname{enumi}{}{} 

\crefname{icase}{case}{cases}
\crefname{ipart}{part}{parts}
\crefname{iprop}{property}{properties}
\crefname{iinv}{invariant}{invariants}

\crefformat{section}{\S\,#2#1#3}
\crefformat{subsection}{\S\,#2#1#3}
\crefrangeformat{section}{\S\S\,#3#1#4--#5#2#6}
\crefmultiformat{section}{\S\S\,#2#1#3}{,~#2#1#3}{,~#2#1#3}{,~#2#1#3}

\usepackage{authblk}
\usepackage[normalem]{ulem}

\usepackage{tikz}
\usetikzlibrary{calc}

\newcommand{\fO}{\mathcal{O}}

\renewcommand{\alpha}{\upalpha}

\newcommand{\connected}[1]{\def\temp{#1}\ifx\temp\empty\sim\else\overset{#1}{\sim}\fi}

\newtheorem{theorem}{Theorem}

\theoremstyle{definition}

\setparsizes{0pt}{0.25\baselineskip plus .25\baselineskip}{1em plus 1fil}

\author[]{L\'{a}szl\'{o} Kozma}

\affil[]{Dresden University of Technology}

\title{Sorting by block-reversals: closing the gap}

\date{}

\begin{document}

\maketitle

\vspace{-0.4in}
\begin{abstract}
In \emph{block-reversal sorting} an elementary operation is to reverse the order of a contiguous block in a sequence, at cost proportional to the length of the block. In SODA 2004, Bender, Ge, He, Hu, Pinter, Skiena, and Swidan showed that every sequence of length $n$ can be sorted at cost $\fO(n\log^2{n})$ and some sequences can only be sorted at cost $\Omega(n\log{n})$. 

In \emph{block-partition sorting}, an elementary operation is to partition the sequence into contiguous blocks, and simultaneously reverse all blocks. The model was introduced in SODA 2024 by Jel\'{i}nek, Opler, and Pek\'{a}rek, who showed that every sequence can be sorted in $\fO(\log^2{n})$ steps and some sequences require $\Omega(\log{n})$ steps. 

We close both gaps, showing that the upper bounds are tight, and the $\log^2{n}$ factors are unavoidable.  
\end{abstract}

 \textbf{Note:} This is a technical report not intended for conference or journal publication. AI tools were used interactively to derive and check essential parts of the result; the writeup is by the author.

\section{Introduction}\label{sec1}

In \emph{block-reversal sorting}, the goal is to re-arrange an input sequence $S = (x_1, x_2, \dots, x_n)$ into a sorted order by elementary operations that reverse a single contiguous block. 

More precisely, given an intermediate state $S' = (x_{\pi_1}, x_{\pi_2}, \dots, x_{\pi_n})$, a block reversal $\sigma(i,j)$ for $i\leq j$ transforms $S'$ into $S'' = (x_{\pi_1},x_{\pi_{2}}, \dots,x_{\pi_{i-1}}, x_{\pi_j}, x_{\pi_{j-1}}, \dots, x_{\pi_{i+1}}, x_{\pi_{i}}, x_{\pi_{j+1}}, x_{\pi_{j+2}}, \dots, x_{\pi_{n}})$. The cost of this operation is $j-i+1$, i.e., the number of elements in the reversed block.  
We assume $S$ to be a permutation of $[n]$, so the goal is to reach the sequence $(1,2,\dots,n)$ with a minimal total cost. 

For an input permutation $S$, denote by $B(S)$ the minimum cost of sorting $S$, and denote the worst-case cost by $B(n) = \max_{S :|S|=n} {B(S)}$.

Motivated by applications in bioinformatics, the problem was introduced by Pinter and Skiena~\cite{PinterSkiena} who showed that $B(n) \in \fO(n\log^2{n})$. 
In follow-up work, Bender, Ge, He, Hu, Pinter, Skiena, and Swidan~\cite{Bender, Benderj} showed $B(n) \in \Omega(n\log{n})$, leaving a logarithmic gap between the upper and lower bounds that has not been closed in the two decades since. In~\cite{Benderj} other weight models are also considered. For most of these, tight bounds are obtained, leaving the (perhaps most natural) linear cost case open.

In a different line of work, Jel\'{i}nek, Opler, and Pek\'{a}rek~\cite{Jelinek} studied sorting under a \emph{block-partition} model. An operation in this setting transforms a sequence $S' = (x_{\pi_1}, x_{\pi_2}, \dots, x_{\pi_n})$ into $S''$, according to a set of indices $1 \leq i_1<i_2< \cdots< i_k < n$,  by applying the block reversals $\sigma(i_j+1,i_{j+1})$ for all $j =0,\dots,k$, with $i_0 = 0$, and $i_{k+1} = n$. In words, we arbitrarily partition the sequence into contiguous blocks and reverse all blocks simultaneously. We define the cost of such a partitioned reversal operation to be $n$, so $f(n)$ such operations cost $n \cdot f(n)$.  
(Note that elements in singleton blocks stay in place, but the cost of an operation is $n$ even in the presence of singleton blocks.) 

For an input permutation $S$, denote by $P(S)$ the minimum cost of sorting $S$ by partitioned reversals, and denote the worst-case cost by $P(n) = \max_{S:|S|=n} {P(S)}$. 

The problem is motivated in~\cite{Jelinek} with the study of hereditary classes of permutations as sorting operators. The authors show an almost complete characterization of the complexity of sorting with various permutation families. The family of \emph{layered permutations} corresponds exactly to the block-partition model defined above. In~\cite{Jelinek} it is shown that $\fO(\log^2{n})$ operations are sufficient and $\Omega(\log{n})$ are sometimes necessary, leaving a logarithmic gap between the upper and lower bounds.

\paragraph{The complexity of sorting.} It is clear that $B(n) \leq P(n)$, since block-reversals can simulate partitioned block-reversals at no greater cost. Thus, it is sufficient to prove upper bounds for $P(n)$ and lower bounds for $B(n)$. The upper bound $P(n) \in \fO(n\log^2{n})$ is straightforward~\cite{Jelinek}. In $\fO(\log{n})$ steps, we can re-arrange the input so that the smallest $\lfloor n/2 \rfloor$ elements appear in the first $\lfloor n/2 \rfloor$ positions. Then, we can recursively and simultaneously sort the first and second halves, leading to cost $t(n) \leq 2\cdot t(n/2) + \fO(n\log{n}) \in \fO(n\log^2{n})$.

A lower bound $P(n) \in \Omega(n\log{n})$ follows from basic information-theory~\cite{Jelinek}. In~\cite{Benderj}, the bound $B(n) \in \Omega(n\log{n})$ is shown via a potential function argument. 

We show the following lower bound that closes the gap in both models, also implying that the asymptotic worst-case complexities of the problems are the same.

\begin{theorem} $B(n) \in \Omega(n\log^2{n}).$ 
\end{theorem}

We assume $n=m^2$, where $m$ is of the form $2^k$. (The result extends to arbitrary $n$ via padding.)

The hard input is the permutation sequence  $S = (x_1, x_2, \dots, x_n)$ defined as 
$$  x_i=m\bigl((i-1)\bmod m\bigr)+\lceil i/m\rceil.$$

It can be seen as filling in an $m \times m$ matrix increasingly row-by-row and reading out the entries column-by-column, 
i.e., $S = (1,\,m+1,\,2m+1,\dots,(m-1)m+1,\,2,\,m+2,\dots,n)$. 

Define the \emph{row-value} of an entry with value $i$ as $row(i) = \lceil i / m \rceil$. In the input sequence, row-values cycle through $1,2,\dots,m$ repeatedly, and in the final sorted sequence they appear in contiguous blocks. 

We define an entropy-based potential that aims to capture how ``mixed'' various contiguous intervals are, in terms of row-values. We show that the total potential must decrease by $\Omega(n\log^2{n})$ while sorting the input, and that the reversal of a single block of length $\ell$ (at cost $\ell$) can only decrease the potential by $\fO(\ell)$. These claims together imply the result.

\paragraph{Potential.} Consider a sequence $S = (x_1, x_2, \dots, x_n)$ with row-values defined as above. Assume that $S$ is padded in both directions with infinitely many virtual elements with row-value $0$.

For a given interval $I$, let $p_{r}(I)$ denote the fraction of elements in $I$ with row-value $r$. The \emph{entropy} of $I$ is then defined as $$H(I) = -\sum_{\substack{0 \leq r \leq m\\p_r >0}} p_r \log{p_r}, \mathrm{~~~~~where~} \log = \log_2.$$ 

The \emph{total entropy} over all intervals of size $w$ (possibly extending beyond the sequence) is:

$$E_w(S) = \sum_i{H(x_i x_{i+1} \cdots x_{i+w-1})}.$$

Note that the sum is over all integers $i$, but only those intervals contribute that overlap with the sequence $S$ (all others consist only of elements with row-value 0, and have zero entropy).

Finally, the potential adds the entropies of intervals of exponentially increasing sizes, up to $m = \sqrt{n}$:

$$\Phi(S) = \sum_{j=1}^k{E_{2^j}}(S).$$

Let $\Phi_{\mathrm{init}}$ denote the potential of the input sequence $S$ defined above, and let $\Phi_{\mathrm{final}}$ be the potential of the sorted sequence. 

In the input every interval $I$ of length $w \leq m$ lying entirely within the sequence has $w$ distinct row-values, and thus, entropy $H(I) =\log w$. There are
$n - w + 1 \geq n - m + 1$ such intervals. 

The initial potential is thus $$\Phi_{\mathrm{init}} \geq (n-m+1) \sum_{j=1}^k \log{2^j} = (n-m+1) \frac{k(k+1)}{2} \in \Omega(n\log^2{n}).$$

When the sequence is sorted, an interval $I$ of length $w \leq m$ has at most two distinct row-values, so its entropy is $H(I) \leq 1$. Only intervals that cross one of the $m + 1$ boundaries between row-values contribute a nonzero term (this includes the two boundaries of the sequence itself). Each boundary lies in $w-1$ intervals of size $w$. 
It follows that 
$$ \Phi_{\mathrm{final}} \leq (m+1) \sum_{j=1}^k (2^j-1) \leq (m+1) \cdot 2m \in \fO(n).$$

It follows that the potential-decrease during the sorting sequence is at least $\Phi_{\mathrm{init}} - \Phi_{\mathrm{final}} \in \Omega(n\log^2{n})$.

\paragraph{Change in intervals due to a block-reversal.} 
It remains to show that reversing a block of length $\ell$ decreases the potential by at most $c \cdot \ell$ for some absolute constant $c$.

Consider some interval $I$ of length $w$, and suppose we reverse some block $B$ between indices $a$ and $b$, of length $\ell = b - a + 1$. Intervals  fully containing block $B$ have their elements permuted, without affecting the entropy of the interval.  Intervals disjoint from $B$ are unaffected. Intervals inside block $B$ are permuted, in the sense that elements of each interval inside $B$ end up as the elements of exactly one interval inside $B$ (possibly the same), so their total entropy is unchanged. 

The only relevant intervals are therefore those that cross exactly one end of $B$. Consider the intervals of this kind that have $q$ positions inside $B$, with
$1 \leq q \leq \min\{w, \ell\}-1$. For each value $q$ there are exactly two such intervals, which we denote $I_L = [a - (w - q), a + q - 1]$
and
$I_R = [b - q + 1, b + (w - q)]$, understanding that the definition is with respect to block $B$ and value $q$. 

\begin{figure}[htbp]
  \centering
  \includegraphics[width=0.6\textwidth]{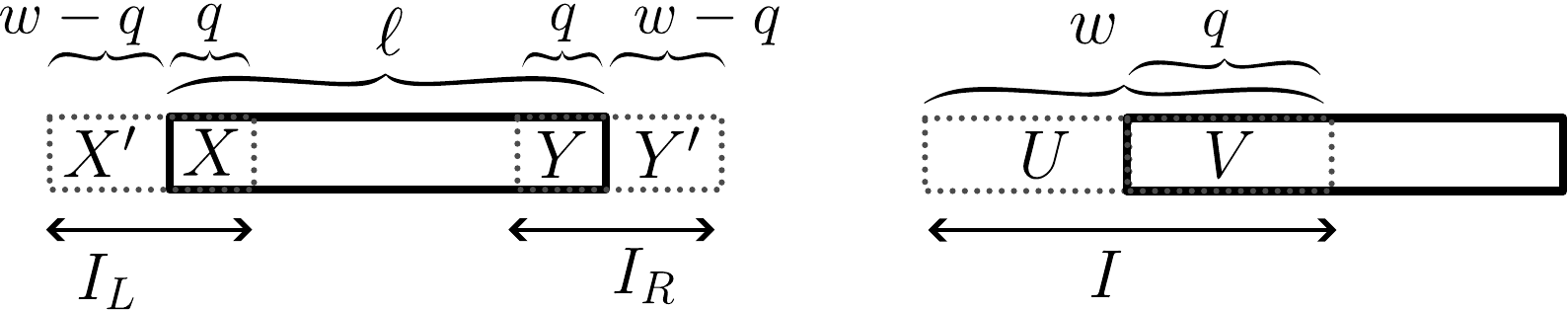}
  \caption{Illustration of the proof. Block $B$ to be reversed shown in bold. }
  \label{fig:example}
\end{figure}

Let $X,Y$ be the length-$q$ prefix and suffix of $B$ and let $X', Y'$ be the $w-q$ positions immediately before
and after $B$, respectively. (See Figure~\ref{fig:example} for illustration.)

Before the reversal we thus have $X'\cdot X$ in $I_L$ and $Y \cdot Y'$ in $I_R$, where $\cdot$ indicates concatenation. After the reversal we have $X' \cdot Y^{\mathrm{rev}}$ in $I_L$, and $X^{\mathrm{rev}} \cdot Y'$ in $I_R$, where $(\cdot)^{\mathrm{rev}}$ indicates 
reversing a block. Note that this holds even if $2q > \ell$, in which case $X$ and $Y$ (and $I_L$ and $I_R$) overlap. 

The important observation is that by flipping $B$, interval $I_L$ loses the elements in $X$ and gains the elements in $Y$, while $I_R$ loses the elements in $Y$ and gains the elements in $X$.   

\paragraph{Bounding the change in entropy.} 
Let $U$ and $V$ be the parts of interval $I$ that are outside and inside the reversed block $B$, with $|V|=q$. If a row-value $r$ occurs $n_r$ times in $I$ and $s_r$ times in $V$, then it occurs $n_r-s_r$ times in $U$. 
By the definition of the entropy, $H(I) = \log{w} - (1/w) \cdot \sum_r n_r \log n_r$. 

Let $p=q/w$ and 
define $h(x)=-x\log x-(1-x)\log(1-x)$, the binary entropy, letting $0\log 0=0$.

Writing the above entropy-formula for $U$ and $V$ and collecting the terms we get  
\begin{align}
\label{eq1}
H(I) = (1-p)H(U)+pH(V)
+ h(p)-\sum_{r:n_r>0}\frac{n_r}{w}\,
h\!\left(\frac{s_r}{n_r}\right).
\end{align}

The weights $n_r/w$ sum to one, and the weighted mean of the ratios $s_r/n_r$ is $\sum_r s_r/w=q/w=p$. Thus, by concavity of $h$, the last sum of~\eqref{eq1} is at most $h(p)$. it follows that the entropy of interval $I$ equals the weighted sum of the entropies of its parts $U$ and $V$ plus a term in $[0,h(p)]$.

Now we consider all intervals of length $w$ that cross exactly one end of $B$. Among these, for all $1 \le q \leq \min\{\ell,w\}-1$, there is exactly one that overlaps with $B$ from the left on $q$ positions, and exactly one that overlaps with $B$ from the right on $q$ positions. Call these $I_L$ and $I_R$. 

As discussed before, after reversing $B$, interval $I_L$ gains exactly the elements of $I_R$ that overlapped with $B$ and $I_R$ gains the elements of $I_L$ that overlapped with $B$, where the gain of one is the loss of the other. Since $I_L$ and $I_R$ have the same $p$, the changes in the $pH(V)$ terms in~\eqref{eq1} for these two intervals cancel out, while the $(1-p)H(U)$ terms remain unchanged. (The reversal of the overlaps with $B$ is irrelevant since the entropy only depends on the counts of row-values.)

Hence, the only change in $H(I_L)+H(I_R)$ is due to the last two terms of~\eqref{eq1}, bounded by $2h(p)$.  

Summing over $q$ gives 
\begin{align}\label{eq3}
|E_w(S'') - E_w(S')| \leq 2 \sum_{q=1}^{\min\{w,\ell\}-1}h(q/w).
\end{align}

\paragraph{Summing over all scales.} Denoting by $\Phi$ and $\Phi'$ the potential before and after the reversal of a block of length $\ell$, the change in potential can be bounded by summing~\eqref{eq3} as follows:
\begin{align}\label{eq2}
|\Phi - \Phi'| \le 2 \sum_{q=1}^{\ell-1} \sum_{\substack{w = 2,4,\dots,2^k\\w > q}} h(q/w).
\end{align}

Fix $q$, and let $w_0$ be the smallest power of two greater than $q$. At scale $w = 2^j w_0$, the
argument $x = q/w$ satisfies $0 < x \leq 2^{-j}$. The function $x\log(\mathrm{e}/x)$ is increasing on $(0, 1]$, and
$h(x) \leq x \log(e/x)$ by the inequality $-(1 - x) \ln(1 - x) \leq x$. Therefore
$$\sum_{\substack{w = 2,4,\dots,2^k\\w > q}} h(q/w) \le \sum_{j=0}^{\infty}{2^{-j}(j+\log{\mathrm{e}})} = 2 + 2\log{\mathrm{e}}.$$

The inner sum in \eqref{eq2} is thus bounded by a fixed constant, and therefore $|\Phi - \Phi'| \leq c \ell$, with $c = 4(1 + \log{\mathrm{e}}) < 10$. It follows that if the reversals have lengths $\ell_1, \dots, \ell_s$ then the total cost is $C = \sum_i{\ell_i}$, hence $C \ge (\Phi_{\mathrm{init}} - \Phi_{\mathrm{final}})/c \in \Omega(n\log^2{n})$. This concludes the proof.

\paragraph{Remark.} While the result settles the worst-case complexity of sorting in the block-reversal and partitioned block-reversal models, relatively little is known about the instance-specific complexity, i.e., of optimally sorting a given input in either model, especially when also taking into account the time spent outside the model, i.e., for deciding which blocks to reverse.

\bigskip

\small
\bibliographystyle{alpha}
\bibliography{main}

\end{document}